\documentclass[
twocolumn, 
groupedaddress, nobibnotes,
prb, byrevtex, 
a4paper, amsfonts]{revtex4}

\usepackage{graphicx}

\begin{document}

\newcommand{\myfigwidth}[0]{26em}

\title{Dinitrosyl formation as an intermediate stage of the reduction
of NO in the presence of MoO$_3$}

\author{Ioannis N. Remediakis$^*$} 
\affiliation{Department of Physics and Division of Engineering and
Applied Sciences, Harvard University, Cambridge MA 02138}
\altaffiliation[Present address: ]{Center for Atomic-scale Materials
Physics, Department of Physics, Building 307, Technical University of 
Denmark, 2800 Lyngby, Denmark}

\author{Efthimios Kaxiras}
\affiliation{Department of Physics and Division of Engineering and
Applied Sciences, Harvard University, Cambridge MA 02138}

\author{Melvin Chen}
\affiliation{Department of Chemistry, Harvard University, Cambridge MA 02138}

\author{Cynthia M. Friend}
\affiliation{Department of Chemistry, Harvard University, Cambridge MA 02138}

\date{\today}

\begin{abstract}
  We present first-principles calculations in the framework of
  density-functional theory and the pseudopotential approach, aiming
  to model the intermediate stages of the reduction of NO in the
  presence of MoO$_3$(010). In particular, we study the formation of
  dinitrosyl, which proves to be an important intermediate stage in
  the catalytic reduction. We find that the replacement of an oxygen
  of MoO$_3$ by NO is energetically favorable, and that the system
  lowers further its energy by the formation of (NO)$_2$. Moreover,
  the geometry and charge distribution for the adsorbed dinitrosyl
  indicates a metal-oxide mediated coupling between the two nitrogen
  and the two oxygen atoms. We discuss the mechanisms for the
  dinitrosyl formation and the role of the oxide in the reaction.

[To be published in the Journal of Chemical Physics]
\end{abstract}

\maketitle



Reduction of nitric oxide (NO) has long been the focus of detailed
studies, because of the role of NO as pollutant in the
atmosphere.  Three-way catalysts, based on Rh, Pt or PdO are used in
automobile exhausts for this purpose; the effectiveness of such
catalysts was found to increase when molybdenum oxides are added
\cite{halasz93,yao84}. Recently, it was proposed that the NO reduction
products N$_2$ and O$_2$ are formed nondissociatevely from NO via an
adsorbed dinitrosyl species, which facilitates N-N coupling
\cite{queeney97}, since low temperature NO coupling proceeds through a
dinitrosyl intermediate \cite{queeney98}. The geometry of the adsorbed
dinitrosyl and the mechanism of N-N coupling remain unknown.  To
address this important question, we perform first-principles
calculations based on density-functional theory and the
pseudopotential approach, for various configurations involving
nitrosyls and the MoO$_3$(010) surface.  The paper is organized as
follows: First, we review the structure of MoO$_3$ and describe the
method we use for the calculations. Next, we present our results for
the MoO$_3$(010) surface with an O vacancy, a structure with an
exposed Mo atom which can adsorb NO molecules. This is followed by
detailed analysis of the configurations involving adsorbed NO and
(NO)$_2$. Finally, we discuss the dinitrosyl formation mechanisms and
the role of the MoO$_3$(010) surface as a catalyst for the reduction
of NO.

\section{Molybdenum trioxide}
\label{sec:moo3}

\begin{figure}
\centering\includegraphics[width=\myfigwidth]{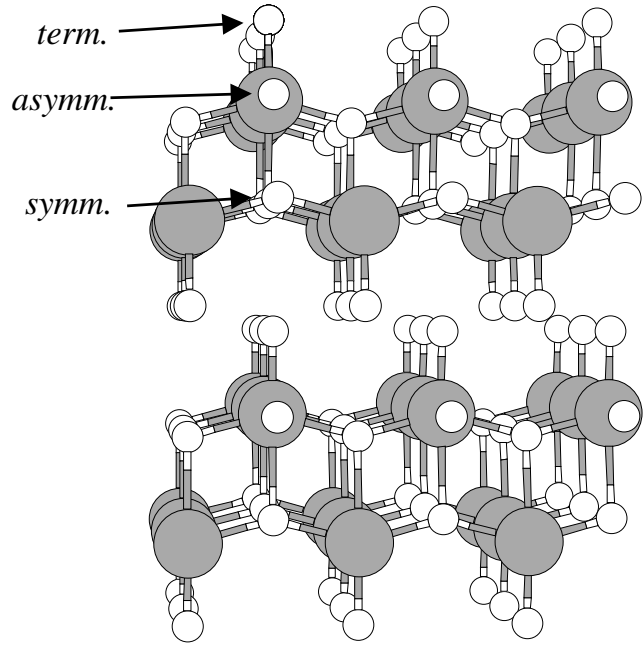}
\caption{Structure of bulk MoO$3$. Mo atoms are shown as larger gray
spheres, O atoms as white, smaller spheres. The three different types
of O atoms (terminal, asymmetric bridging, symmetric bridging) are
indicated.}
\label{fig:moo3bulk}
\end{figure}

Molybdenum trioxide is a layered material, its layers being weakly
bonded via van der Waals interactions. The space group is $Q_h^{16}$
$(bnm)$, and the lattice is orthorhombic \cite{wooster31} with
parameters \cite{wyckoff} 3.962, 13.858, and 3.697 \AA.  Each unit
cell contains four MoO$_3$ units. The lattice coordinates of the atoms
are $\pm(x, y, \frac{1}{4})$ and $\pm(\frac{1}{2}-x, \frac{1}{2}+y,
\frac{1}{4})$, with ($x$, $y$) equal to (0.086, 0.099) for the Mo
atoms , and (0.086, 0.250), (0.586, 0.099), (0.586, 0.431) for the
three O atoms surrounding each Mo atom, respectively\cite{wyckoff}.

The MoO$_3$ crystal is shown in Fig. 1. Macroscopically, the material
forms ``small, thin, lustrous plates, parallel to (010)''
\cite{wooster31}. This is revealed in the atomistic structure: It
consists of bilayers parallel to the (010) plane, which are bonded
through weak electrostatic interactions, with the (010) surface being
the easy cleavage plane of the crystal. Each bilayer consists of two
sublayers of periodically arranged distorted MoO$_6$ octahedra. There
are three structurally different O atoms. The {\em asymmetric
bridging} oxygen is collinear with two Mo atoms and forms one long and
one short bond with them. The {\em symmetric bridging} oxygen is
located between the two sublayers of the bilayer and bonds to two Mo
atoms of one sublayer with equal bond lengths and to one Mo atom of
the other sublayer with an elongated bond.  Finally, the univalent
{\em terminal} oxygen is connected to one Mo atom forming the shortest
Mo-O bond in the system. The Mo-O bond lengths in MoO$_3$ as obtained
from several experimental and theoretical studies are shown in Table
\ref{tab:bonds}.

\begin{table*}
\caption{Bond lengths (in \AA) between Mo and O in MoO$_3$ as
calculated in the present work compared to other calculations (a) by
Chen {\em et al.} \cite{chen98} and (b) by Yin {\em et al.}
\cite{yin01} and to experimental data obtained from (c) the atomic
positions given by Wyckoff \cite{wyckoff} and and (d) the work of
Kihlborg \cite{kihlborg63} as given in Ref. \cite{chen98}. The symbols
O$_t$, O$_s$ and O$_a$ denote the terminal, symmetric bridging and
asymmetric bridging O atoms respectively.} 
\begin{center}
\begin{tabular} {|c|r|rr|rrr|} \hline
Bond & \multicolumn{6}{c|}{Bond Length} \\ \hline
  & This work & Theory (a) & Theory (b) & Exp. (c) & Exp. (d) & Exp. (avg)\\ \hline
Mo-O$_t$ & 1.77      & 1.67  &  1.76            & 1.82  & 1.67  & 1.75 \\ 
Mo-O$_s$ & 1.96      & 1.92  &  2.02            & 1.93  & 1.95  & 1.94 \\
         & 2.33      & 2.30  &  2.28            & 2.32  & 2.33  & 2.33 \\
Mo-O$_t$ & 1.87      & 1.76  &  1.81            & 1.89  & 1.73  & 1.81 \\
         & 2.17      & 2.19  &  2.25            & 2.08  & 2.25  & 2.17 \\ 
\hline
\end{tabular}
\end{center}
\label{tab:bonds}
\end{table*}


\section{The calculation}

All calculations reported in this work were performed via the
High-performance-Fortran Adaptive Real-space Electronic Structure
(HARES) package \cite{modine97}. We use Density-Functional Theory
(DFT) in the Local Spin Density Approximation (LSDA)
\cite{hohenberg64,kohn65}. The Kohn-Sham valence electron
wavefunctions are represented in a real-space orthogonal grid with a
spacing of 0.13 \AA, which is equivalent to a cut-off energy of 150 Ry
in a plane-wave basis calculation
The ionic cores and their interaction with valence electrons were
taken into account through the soft Troullier-Martins
\cite{troullier91} pseudopotentials, in the separable
Kleinman-Bylander form \cite{kleinman82}. The Laplacian is expanded to
second order terms. For the exchange-correlation functional we use the
results of Ceperley and Alder as parametrized by Perdew and Zunger
\cite{ceperley80}. The electronic density was updated during the
self-consistent loop using a modified Broyden mixer \cite{jain02}.

We model the surface of MoO$_3$ by a single bilayer perpendicular to
the (010) direction. This approximation is not expected to affect the
results, due to the weak inter-layer coupling. The calculated Mo-O
bond lengths for the single-bilayer system are shown in Table
\ref{tab:bonds}. As experimental data have discrepancies, we compare
our results to the average of the two sets (last column of Table
\ref{tab:bonds}). The agreement to experiment is within 2\%. Our
results are also in agreement to the calculation of Yin {\em et al.}
\cite{yin01}.  Moreover, vibrational frequencies calculated within the
single bilayer approximation \cite{chen98} are also in very good
agreement with experiment. This justifies our choice to use a
simplified unit cell, while establishing the adequacy of DFT and HARES
to correctly describe bonding in the material under study.  The
surface unit cell we chose has $(2\times 2)$ periodicity relative to
the ideal structure. This is sufficient to simulate isolated O
vacancies or adsorbed NO, due to the local character of the
substrate-adsorbate interaction and the lack of reconstruction effects
for this surface. The Brillouin Zone was sampled by the $\Gamma$ point
only, which is adequate due to the large size of the unit cell. The
atomic degrees of freedom were relaxed with the
Broyden-Fletcher-Goldfarb-Shanno (BFGS) method \cite{chetty95,shano78}
until the calculated Hellman-Feynman forces were smaller in magnitude
than 0.005 a.u. We used the symmetry of the problem to eliminate some
electronic as well as ionic degrees of freedom where appropriate (see
below). We have checked these computational parameters against
previous calculations \cite{modine97,waghmare01} of similar systems to
make sure they give well-converged results.


\section{Terminal O vacancy and NO adsorption}
\label{sec:moo3v}

\begin{figure}
\centering\includegraphics[clip=true, bb=169 74 845 685,
  width=\myfigwidth]{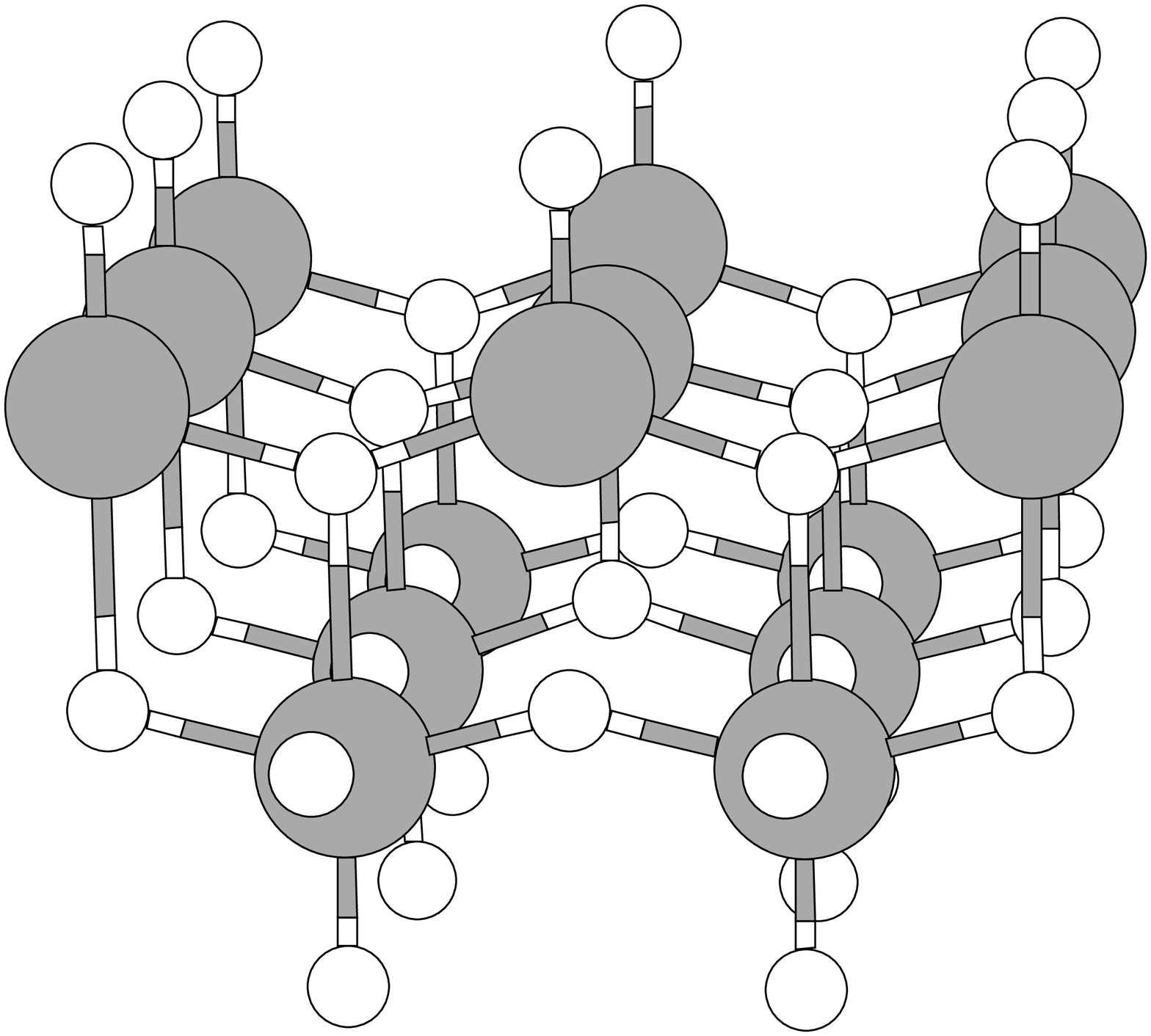} {\Large (a)}

\centering\includegraphics[clip=true, bb=169 74 845 695,
  width=\myfigwidth]{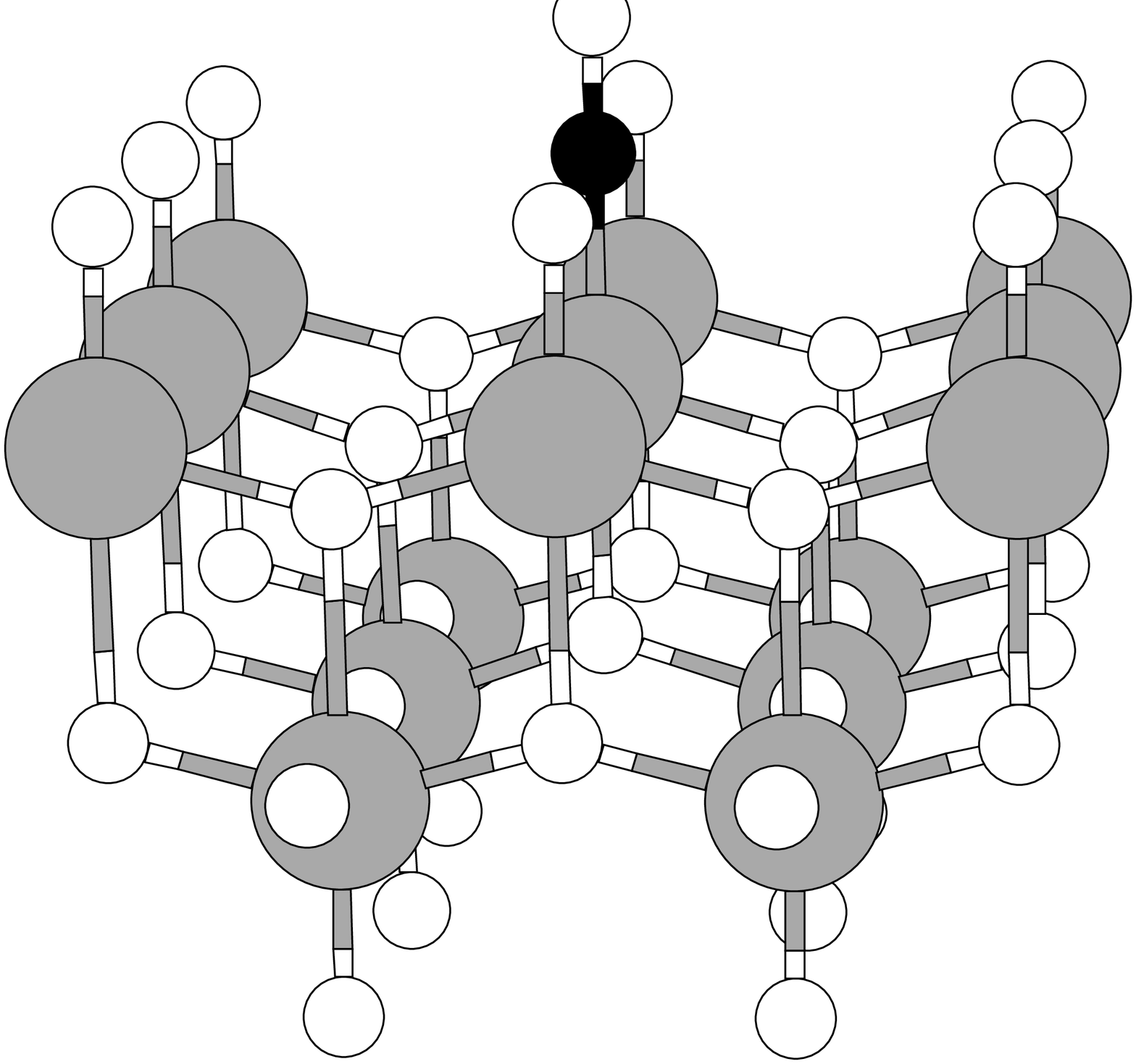} {\Large (b)}
\caption{Relaxed geometries for MoO$_3$(010) (a) with a terminal O vacancy
and (b) with an adsorbed NO molecule. Grey, white and black spheres
correspond to Mo, O and N atoms, respectively. }
\label{fig:moo3v}
\end{figure}

An O vacancy in MoO$_3$ is formed when a terminal oxygen atom is missing. As
discussed in Section \ref{sec:moo3}, the terminal O is univalent and forms the
shortest Mo-O bond in the bulk; this bond is also the strongest, with a
calculated \cite{mihalak96} bond order of 1.93. Accordingly, it is expected
that the formation energy of the vacancy will be high. Indeed, our calculation
yields a formation energy of 2.95 eV/vacancy, assuming that the O atoms that
leave the surface form O$_2$ molecules (see Table \ref{tab:reactions}). The
$(2\times 2)$ unit cell corresponds to a 0.25 ML coverage of vacancies, which
is low enough to ensure that the interactions between vacancies are
negligible.  The formation of the O vacancy is accompanied by small
relaxations of atoms around the missing terminal O atom. The relaxed geometry
is shown in Fig. 2(a). The main effect is an inward displacement
of the exposed Mo atom, which results in shrinking the Mo-symmetric oxygen
bonds from 1.96 \AA~to 1.91 \AA~for O in the same sublayer and from 2.33
\AA~to 2.11 \AA~for O in the adjacent sublayer.  None of the atoms has any
significant displacement parallel to the (010) plane; we thus find no evidence
of surface reconstruction due to vacancy formation.

The MoO$_3$(010) surface with a terminal oxygen vacancy, shown in
Fig. 2(a), has many characteristics of a good catalytic
platform: The exposed Mo atom promises to be chemically active and has
the ability to bond to more than one adsorbate atoms, since it has
lost a double bond. The remaining terminal O atoms around the vacancy
site form a cage structure that can enhance coupling between the
adsorbates. In ideal MoO$_3$, the calculated \cite{mihalak96}
population for a terminal O is -0.4 electrons, so the remaining
terminal O atoms are expected to repel the adsorbates, creating an
implicit attraction between them.

The first step in the catalytic reduction of NO is the adsorption of a NO
molecule. The binding configuration is shown in Fig. 2(b). NO
is bonded to a Mo atom and substitutes a terminal O. The molecule is
perpendicular to the surface in the lowest energy geometry, rendering the
dissociation of nitrosyl on the surface implausible. The adsorption of NO on
MoO$_3$ with an O vacancy restores the positions of the surrounding atoms to
their positions in the ideal MoO$_3$(010) surface: the lengths of the Mo-N and
N-O bonds are 1.91 and 1.14 \AA, respectively.  The N-O bond has very similar
length to the bond of a free NO molecule. We find this bond to be 1.15
\AA, in excellent agreement to experiment.

The binding energy for NO to the MoO$_3$(010) surface with an oxygen
vacancy is 2.98 eV/molecule. The energy difference between the
MoO$_3$(010) surface and the same surface with NO substituting a
terminal O is $-$0.03 eV/molecule, assuming that the O atoms leaving
the surface form O$_2$ molecules. The system lowers its total energy
by the substitution of terminal O by NO, so this configuration must be
an important step in the NO reduction. The mechanism of the
substitution and the formation of O$_2$ gas is not known, but it seems
unlikely that the system passes through the O vacancy configuration,
since the barrier for this path would be at least 2.95 eV, a number
which would forbid the process under ordinary conditions. The
transition state for the substitution could involve intermediate
formation of NO$_2$ and/or O$_3$.

NO adsorption could also take place after a vacancy on the surface has
been formed by an independent mechanism, for example after a CO
molecule has hit the surface and CO$_2$ has been formed. Another
similar mechanism would be the formation of H$_2$O from gas-phase
H$_2$ and a terminal O atom. The exact final product of the reduction
of MoO$_3$(010) to MoO$_3$[v](010) cannot be {\em ad hoc} determined,
as it depends on the experimental conditions. However, the final state
of the O atom that leaves the surface to create the vacancy would not
affect the picture, as the energetics would not change
dramatically. We therefore choose to use O$_2$ as a typical product of
a reduction reaction.

\section{Dinitrosyl chemisorption}
\label{sec:moo3_2no}

\begin{figure}
\centering\includegraphics[clip=true, bb=169 74 845 685,
  width=\myfigwidth]{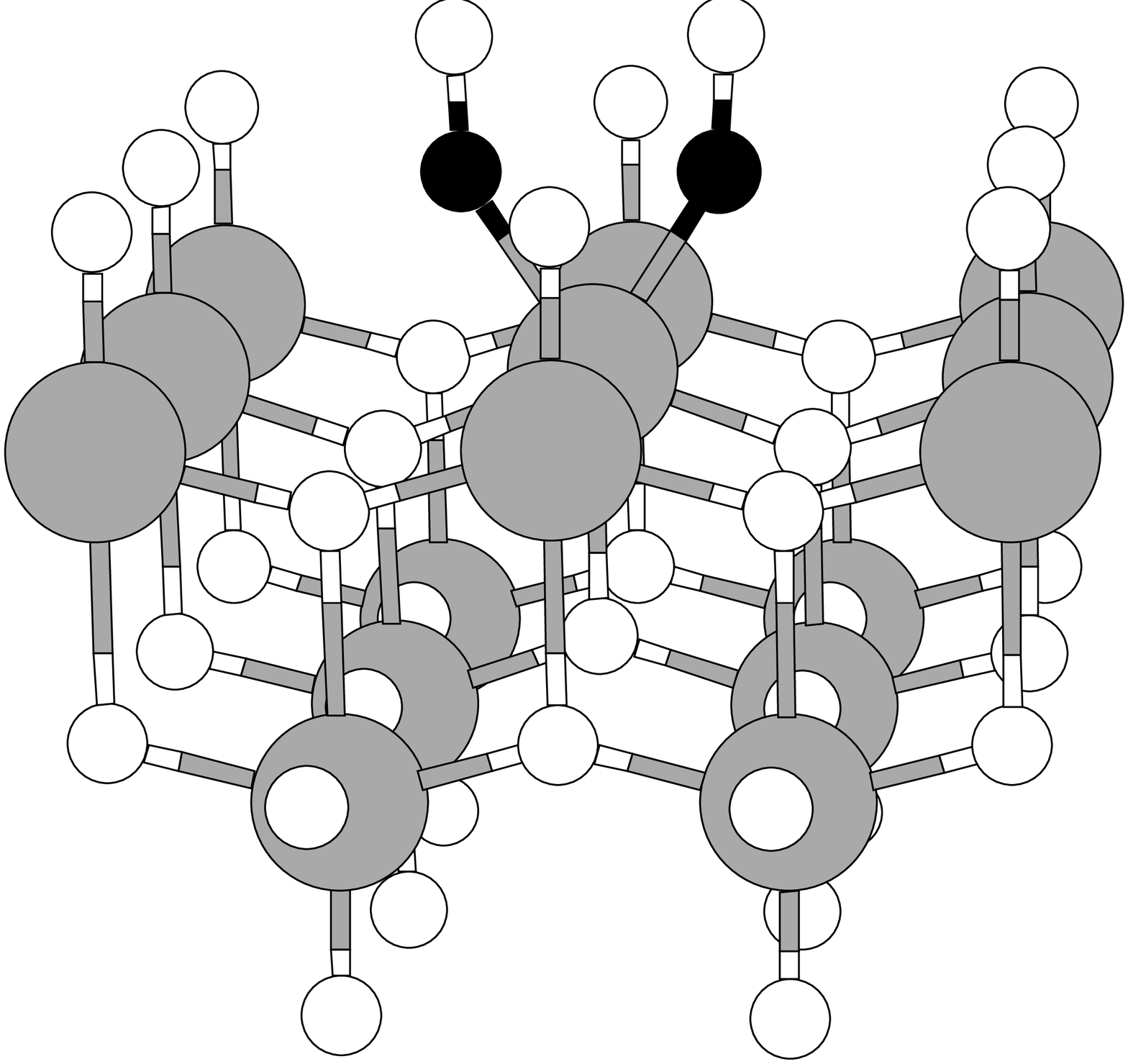} {\Large (a)}
\centering\includegraphics[clip=true, bb=169 74 845 695,
  width=\myfigwidth]{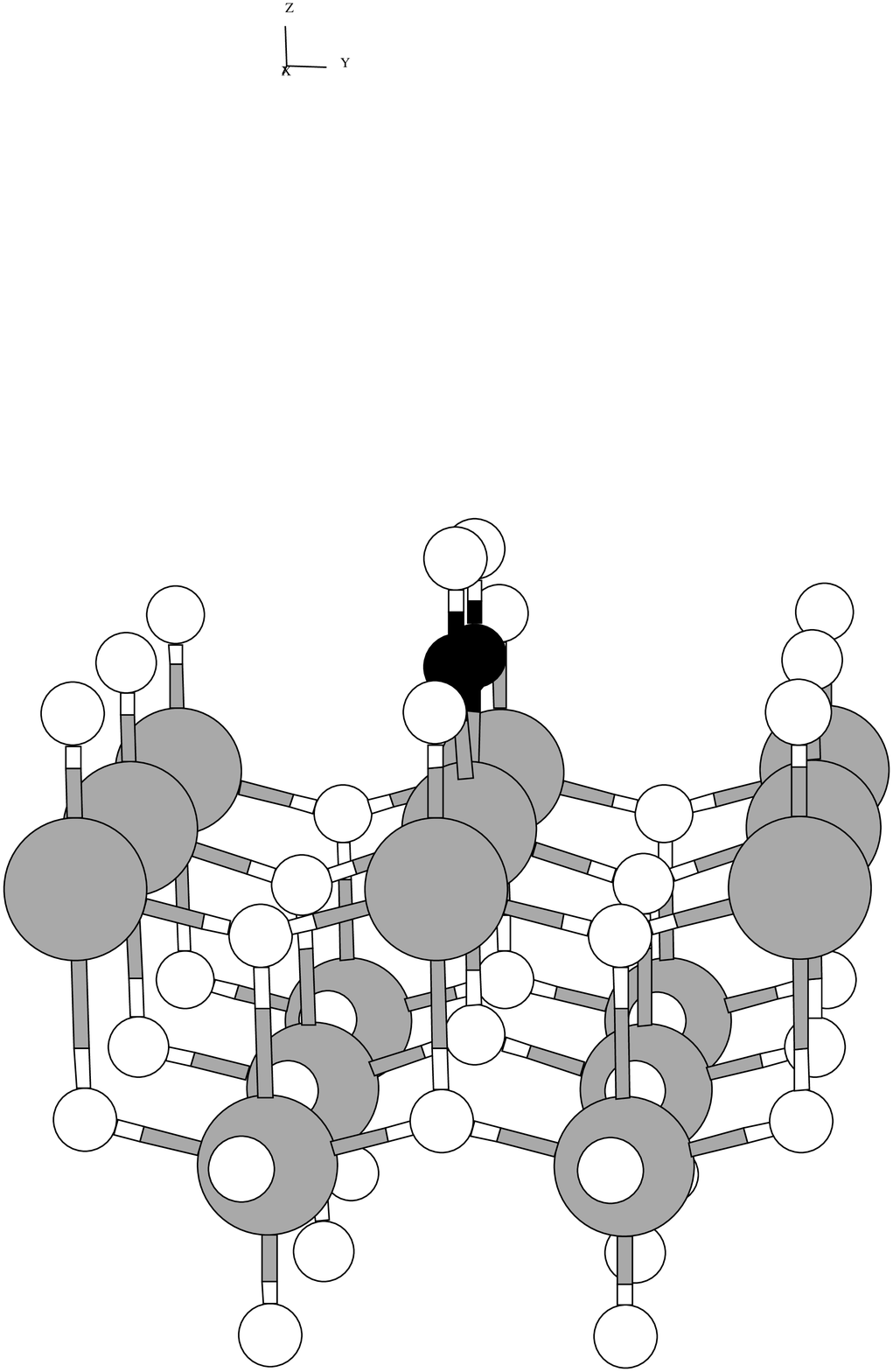} {\Large (b)}
\caption{Relaxed geometries for dinitrosyl adsorption (a) in the 
symmetric O plane and (b) in the asymmetric O plane. }
\label{fig:moo3_2no}
\end{figure}

To study the energetics of the dinitrosyl adsorption, we consider two
characteristic configurations: In the first, the (NO)$_2$ plane is
parallel to (001) or the plane defined by the Mo atom and the
symmetric bridging O atoms. In the second configuration, the (NO)$_2$
plane is parallel to (100), or a plane perpendicular to the previous
one which contains the Mo atom and the asymmetric bridging O atoms.
The relaxed geometries for both configurations are shown in Fig. 3.

In the first case, a reflection symmetry with respect to the (100)
plane was imposed, following the experimental results obtained on
oxidi\-zed Mo(110) \cite{queeney97}. Removal of the reflection
symmetry does not result in further relaxation or significant lowering
of the total energy, indicating that in the lowest energy geometry the
two NO molecules of the dinitrosyl are identical. By contrast,
(NO)$_2$ adsorbed parallel to the (100) plane is found to be strongly
asymmetric, with one of the two NO molecules closer to the Mo center
than the other. This is a direct consequence of the asymmetry of the
underlying oxygen atoms.

\begin{table}
\caption{Bond lengths (in \AA) for the various structures discussed in
the text. O$_t$, O$_s$ and O$_a$ stands for the terminal, symmetric
bridging and asymmetric bridging O, respectively. The Mo-O$_s$ bonds
for Mo and O$_s$ belonging at the same or adjacent sublayers, as well
as the short and long Mo-O$_a$ bond are shown. (s) and (a) indexes for
adsorbed (NO)$_2$ refer to adsorption with the molecule parallel to
the symmetric and asymmetric bridging O planes, respectively.}

\begin{center}
\begin{tabular}{|c|c|c|c|c|c|c|c|c|} \hline 
Structure &    Mo - O$_t$ & \multicolumn{2}{|c|}{Mo-O$_s$}
& \multicolumn{2}{|c|}{Mo-O$_a$} & Mo-N & N-O & N-N \\ \hline
                &       & same & adj. & short& long &      &      &\\
\hline
Ideal           & 1.77  & 1.96 & 2.33 & 1.87 & 2.17 & -    & -    & - \\
O vacancy       & 1.77  & 1.91 & 2.11 & 1.90 & 2.06 & -    & -    & - \\
ads. NO         & 1.77  & 1.95 & 2.23 & 1.86 & 2.15 & 1.91 & 1.14 & - \\
ads. (NO)$_2$(s)& 1.78  & 1.98 & 2.36 & 1.89 & 2.16 & 2.00 & 1.14 & 2.18\\
ads. (NO)$_2$(a)& 1.78  & 1.96 & 2.29 & 1.90 & 2.18 & 1.95 & 1.15 & 2.21\\
                &       &      &      &      &      & 2.35 & 1.13 & \\
\hline
\end{tabular}
\end{center}
\label{tab:allbonds}
\end{table}

The bond lengths for both configurations, together with the bond
lengths of the previously discussed structures are summarized in Table
\ref{tab:allbonds}. For adsorbed dinitrosyl in the Mo-symmetric O
plane, the Mo-N bond is 2.00 \AA~ long, 5\% longer than in that of a
single adsorbed NO. The Mo-N bond is weaker for adsorbed (NO)$_2$ as
the bonding electrons of Mo are shared by two N atoms. The N-O bond
length is 1.14 \AA, identical to that of adsorbed NO. In the second
configuration, having dinitrosyl in the Mo-asymmetric O plane, the
Mo-N (N-O) bonds have lengths 1.95(1.15) and 2.35(1.13) \AA, for NO
above the long (2.18 \AA) and short (1.90 \AA) Mo-asymmetric O bond,
respectively. Both N and O of the adsorbed nitrosyls are more strongly
bound above the less strongly bonded asymmetric O.

The asymmetric configuration, with (NO)$_2$ parallel to the
Mo-asymmetric O plane, has a slightly lower energy than the other, by
0.10 eV per (NO)$_2$ molecule. The binding energy to MoO$_3$(010) with
an oxygen vacancy is 3.80 eV and 3.90 eV per molecule for the
symmetric and asymmetric configuration, respectively. The energy gain
when (NO)$_2$ substitutes a terminal O of MoO$_3$(010) with formation
of O$_2$ gas is 0.85 eV and 0.95 eV per molecule for the two
configurations.

\begin{figure}
\centering\includegraphics[width=\myfigwidth]{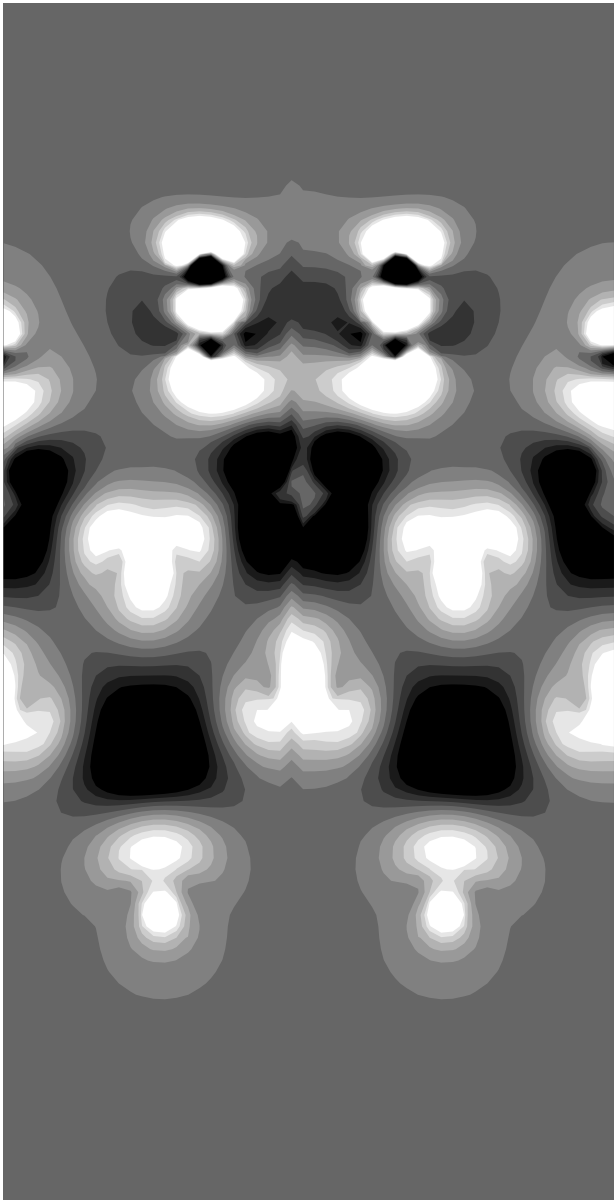} 
\caption{Contour plot of total electronic density minus a 
  superposition of valence atomic charge densities in the Mo-N-O plane
  for adsorbed (NO)$_2$ parallel to the symmetric O plane. The
  background shade corresponds to zero; darker shades correspond to
  negative values, lighter shades to positive values.}
\label{fig:density.diff}
\end{figure}

The large lowering of the total energy by 0.90 eV on average per
molecule for (NO)$_2$ adsorption, compared to 0.03 eV per NO for the
NO adsorption, indicates that bonding between the two parts of the
dinitrosyl has taken place. This bonding is evident in Fig. 4, which
presents a contour plot of the total electronic density minus a
superposition of atomic electronic densities on the dinitrosyl plane,
which coincides with the symmetric bridging O plane. According to our
choice of shading, electronic charge has moved from the darker to the
lighter regions, relative to the charge distribution of a
superposition of atomic densities. The white T-shaped clouds inside
the slab correspond to symmetric bridging O atoms, while those at the
bottom to terminal O atoms. Small gray spots in the middle of black
regions correspond to Mo atoms. The two bright vertical complexes
correspond to the two NO molecules. As the electronegativities of the
involved elements dictate, charge has moved from the Mo atoms to N and
O atoms, with O atoms having slightly more charge concentrated around
them compared to N atoms.

We observe an apparent electron sharing between the two N atoms,
manifesting the N-N coupling due to the presence of the catalytic
surface. A similar, but weaker, coupling exists between the O atoms of
the dinitrosyl. The absence of bonding charge between the N atoms of
the adsorbate and the surrounding terminal O atoms implies a repulsion
between them. It is this repulsion that leads to the attraction
between the N atoms.

\section{Discussion}

\begin{table*}
\caption{Energetics of the reactions discussed in the text, as
calculated in the present work. MoO$_3$(010)[v] refers to a
MoO$_3$(010) surface with a terminal O vacancy; (g) stands for
gas-phase molecule and (a) for adsorbed molecule. The energies for 
the reactions involving adsorbed dinitrosyl are the averages of the two
considered configurations mentioned in Section \ref{sec:moo3_2no}.}
\begin{center}
\underline{Vacancy formation:} \\
MoO$_3$(010) $\rightarrow$ MoO$_3$(010)[v] $+$ $\frac{1}{2}$O$_2$(g)
$-$2.95 eV

\underline{NO adsorption:} \\
MoO$_3$(010)[v] $+$ NO(g) $\rightarrow$ [MoO$_3$(010)[v] $+$ NO(a)] $+$ 2.98 eV \\
MoO$_3$(010) $+$ NO(g) $\rightarrow$ [MoO$_3$(010)[v] $+$ NO(a)] 
   $+$ $\frac{1}{2}$O$_2$(g) $+$0.03 eV

\underline{(NO)$_2$ adsorption:} \\
MoO$_3$(010)[v] $+$ 2NO(g) $\rightarrow$ 
      [MoO$_3$(010)[v] $+$ (NO)$_2$(a)]  $+$ 3.85 eV \\
MoO$_3$(010) $+$ 2NO(g) $\rightarrow$ 
      [MoO$_3$(010)[v] $+$ (NO)$_2$(a)]  $+$ $\frac{1}{2}$O$_2$ $+$ 0.90 eV

\underline{(NO)$_2$ formation from adsorbed NO:} \\
2[MoO$_3$(010)[v] $+$ NO(a)] $\rightarrow$
       [MoO$_3$(010)[v] $+$ (NO)$_2$(a)] $+$ MoO$_3$(010)[v] $-$2.09 eV \\
2[MoO$_3$(010)[v] $+$ NO(a)] $+$ $\frac{1}{2}$O$_2$(g) $\rightarrow$
       [MoO$_3$(010)[v] $+$ (NO)$_2$(a)] $+$ MoO$_3$(010)
       $+$0.86 eV \\  
~[MoO$_3$(010)[v] $+$ NO(a)] $+$ NO(g) $\rightarrow$ 
       [MoO$_3$(010)[v] $+$ (NO)$_2$(a)] $+$ 0.87 eV \\
\end{center}
\label{tab:reactions}
\end{table*}

The reactions leading to dinitrosyl formation, starting either with gas phase
NO or with adsorbed nitrosyl, are summarized in Table \ref{tab:reactions}.
Although the adsorption of two gas NO molecules to form a dinitrosyl is
exothermic by 0.90 eV/molecule (this number is the average between the two
characteristic geometries of the dinitrosyl discussed before), the process is
expected to have a large energy barrier due to the repulsion of the NO
molecules, which, having an electrical dipole moment, would prefer to have an
anti-parallel configuration. The combination of two adsorbed nitrosyls is also
rejected as a mechanism for dinitrosyl formation, as the vacancy left behind
makes the process energetically costly. Indeed, as shown in Table
\ref{tab:reactions}, this reaction is endothermic by more than 2 eV/molecule.
The alternative is to fill the O vacancy by an O from the environment. In this
case, the dinitrosyl formation is exothermic by 0.81/0.91 eV/molecule. This
process, being the opposite of the NO adsorption, would involve the same
transition state and consequently requires overcoming almost the same energy
barrier.

A lower energy barrier, with almost the same energy gain, is possible
when a gas-phase NO molecule binds to an already adsorbed one (last
reaction of Table \ref{tab:reactions}). The energy barrier for this
case has to be lower than in the previous one, since this process
involves no breaking of bonds, while the terminal oxygens of MoO$_3$
surrounding the adsorbed NO will attract the N atom of the gas-phase
NO. The calculated lowest energy, together with the expected low
barrier, support the presumption that this is the dominant dinitrosyl
formation mechanism.

Combining the previous results, the proposed mechanism for the
reduction of NO is described in the following series of reactions:

\begin{center}
\begin{tabular}{rl}
2NO(g) & $\rightarrow$ NO(g) + NO(a) + $\frac{1}{2}$O$_2$ + 0.03 eV \\
       & $\rightarrow$ (NO)$_2$(a) +  $\frac{1}{2}$O$_2$ + 0.87 eV  \\
       & $\rightarrow$ N$_2$ + O$_2$ + 1.00 eV.
\end{tabular}
\end{center}

A gas NO molecule is exchanged with a MoO$_3$(010) terminal O, and the
system lowers its energy by 0.03 eV.  Alternatively, as mentioned in
Section \ref{sec:moo3v}, a vacancy could be created by the reaction of
the surface with some other gas-phase molecule; in that case the
products of the reaction would not be necessarily include
O$_2$. Another NO molecule binds to the already adsorbed one, lowering
the energy of the system by about another 0.84 eV. The next step is
either desorption of the two O atoms first (as O$_2$) followed by
desorption of the two N atoms (as N$_2$), or desorption of one O atom
first and then of the N$_2$O molecule. In both cases, an O atom fills
the O vacancy that is left behind so that the catalyst is left
unchanged at the end of the process. The final state of the system,
with N$_2$ and O$_2$ molecules is \cite{crc2000} 1.90 eV lower in
energy than the initial state of two NO molecules.  The contribution
of the catalyst in the process is twofold: first, it can reduce the
barrier for NO reduction, and second, it makes it much more probable
for two NO atoms to come close to each other and react.

\begin{acknowledgments}  
  I.R is grateful to Dr. Leeor Kronik for valuable advice regarding
  high-performance Broyden mixing algorithms, and to Dr. Umesh
  Waghmare for stimulating discussions. This work was supported in
  part by Harvard's Materials Research Science and Engineering Center,
  which is funded by the National Science Foundation.
\end{acknowledgments}

\end{document}